\documentclass[format=acmlarge]{acmart}

\def\BibTeX{{\rm B\kern-.05em{\sc i\kern-.025em b}\kern-.08emT\kern-.1667em\lower.7ex\hbox{E}\kern-.125emX}}
    
\copyrightyear{2019}
\acmYear{2019}
\setcopyright{none}
\acmConference[]{}{}
\acmBooktitle{}
\acmPrice{}
\acmDOI{}
\acmISBN{}

\usepackage{subfig}
\usepackage{multirow}
\usepackage{multicol}

\begin{document}

%
% The "title" command has an optional parameter, allowing the author to define a "short title" to be used in page headers.
\title{Real World Longitudinal iOS App Usage Study at Scale}

%
% The "author" command and its associated commands are used to define the authors and their affiliations.
% Of note is the shared affiliation of the first two authors, and the "authornote" and "authornotemark" commands
% used to denote shared contribution to the research.
\author{Dohyun Kim}
\author{Joshua Gluck}
\affiliation{%
  \institution{Carnegie Mellon University}
  \department{School of Computer Science}
  \city{Pittsburgh}
  \state{PA}
  \postcode{15213}
  \country{USA}
}
\author{Malcolm Hall}
\affiliation{%
  \institution{University of Glasgow}
  \city{Glasgow}
  \postcode{G12 8QQ}
  \country{UK}}
\author{Yuvraj Agarwal}
\affiliation{%
  \institution{Carnegie Mellon University}
  \department{School of Computer Science}
  \city{Pittsburgh}
  \state{PA}
  \postcode{15213}
  \country{USA}
}

%
% By default, the full list of authors will be used in the page headers. Often, this list is too long, and will overlap
% other information printed in the page headers. This command allows the author to define a more concise list
% of authors' names for this purpose.
%\renewcommand{\shortauthors}{Kim and Gluck, et al.}

% The abstract is a short summary of the work to be presented in the article.
\begin{abstract}
Given the importance of understanding the interaction between mobile devices and their users, app usage patterns have been studied in various contexts. However, prior work has not fully investigated longitudinal changes to app usage behavior. In this paper, we present a longitudinal, large-scale study of mobile app usage based on a dataset collected from $162,006$ iPhones and iPads over 4 years. We explore multiple dimensions of app usage pattern proving useful insights on how app usage changes over time. Our key findings include (i) app usage pattern changes over time both at the individual app level and the app category level (i.e. proportion of time a user spends using an app), (ii) users keep a small set of apps frequently launched (90\% of iPhone users launch roughly 14-18 apps weekly), (iii) a small number of apps remain popular while some specific kinds of apps (e.g. Games) have a shorter life cycle compared to other apps of different categories. Finally, we discuss our findings and their implications, for example, a short-term study as an attempt to understand the general needs of mobile devices may not achieve useful results for the long term.
\end{abstract}

 \begin{CCSXML}
<ccs2012>
<concept>
<concept_id>10003120.10003121.10011748</concept_id>
<concept_desc>Human-centered computing~Empirical studies in HCI</concept_desc>
<concept_significance>500</concept_significance>
</concept>
</ccs2012>
\end{CCSXML}

\ccsdesc[500]{Human-centered computing~Empirical studies in HCI}
%
% Keywords. The author(s) should pick words that accurately describe the work being
% presented. Separate the keywords with commas.
\keywords{smartphone, mobile application usage, user study}

%
% A "teaser" image appears between the author and affiliation information and the body 
% of the document, and typically spans the page. 
%%  \includegraphics[width=\textwidth]{sampleteaser}
%%  \caption{Seattle Mariners at Spring Training, 2010.}
%%  \Description{Enjoying the baseball game from the third-base seats. Ichiro Suzuki preparing to bat.}

%
% This command processes the author and affiliation and title information and builds
% the first part of the formatted document.
\maketitle

\section{introduction}

% General Introduction - Sketch of Status Quo : Mobile devices are ubiquitous
Over the last decade, starting from the introduction of the iPhone and iOS, the adoption of mobile devices (e.g. smartphones) has been unprecedented, with billions of devices already in use. This growth has been fueled by extremely capable devices with functionality driven by rich sensing modalities, significant computational power, and most importantly millions of apps developed for them. We can now use our phones for diverse tasks well beyond communication: from controlling our smart home appliances, to entertainment and content creation, to health and fitness tracking. Recent surveys (2014) from Bank of America reported that people ranked their smartphones as more important than PCs and that a majority of people could not cope without their phone for even a day \cite{phoneuse}. 

% Importance of understanding usage behavior - Implications of usage behavior analysis : Examples of UX & Recommendation Engines
Given the importance of mobile devices and the app ecosystems around them, understanding app usage modalities is key for reasons ranging from building compelling and better apps to providing an overall improved user experience. For example, iOS Spotlight has a basic suggested app feature that works by monitoring the time and location of previous app use. Similarly, by knowing which apps or content a user might be interested in, the operating system can preload apps to reduce perceived delay \cite{yan_fast_2012}. Additionally, if users with similar app usage patterns, as well as other contextual parameters, can be grouped together they can be recommended new apps that others in their cluster use \cite{xu_preference_2013}.

This second use case is particularly important as there are millions of unique apps in both Apple's App Store and Google's Play Store, making the discovery of new apps incredibly hard for users. Similarly, recent work on managing user privacy on Android devices noted that building user profiles of ``similar'' users can be useful in providing recommendations to the users belonging to the same profile \cite{liu_follow_2016}. Finally, app developers can potentially target or customize their apps for certain user types based on context, for example people using phones only during the day, at night, those who use certain types of apps (e.g. Games), or who use their devices more or less than a certain amount each day. 
However, for many of the use cases highlighted above we ideally need to understand whether app usage changes over time, and what factors affect it, to account for any long-term temporal effects in user behavior.

% Limitation of previous work - Research question remains for temporal perspective
Recognizing the importance of mobile device usage, there has been a growing research interest in examining usage data from mobile devices. These include analyzing how long application sessions last \cite{banovic_proactivetasks:_2014,bohmer_falling_2011}, the relationship between mobile app usage and mobile search functions \cite{carrascal_-situ_2015}, using Markov Models to study and represent device usage \cite{kostakos_modelling_2016}, contextual factors affecting app usage \cite{xu_preference_2013,yan_mobisys11_appjoy,do_ICMI_2011_smartphoneusage, Hintze:2017}
, the factors leading to repeated usage of an app\cite{jones_revisitation_2015}, as well as emerging meta-analysis over the trials and best-practices for studying mobile device usage \cite{church_understanding_2015}. More recently, Zhao et al. analyzed app usage data for Android users over a month to show that user populations are not homogeneous and are in fact comprised of many sub-groups \cite{zhao_discovering_2016}.

% New Contribution
In this paper, we investigate whether \emph{app usage pattern changes over time} using the dataset that covers 4 years of users interacting with apps on their smart devices. While various aspects of app usage patterns have been studied for shorter time periods (e.g. up to a few months), it is not well-known how these app usage patterns change over a more extended period of time (e.g. multiple years). We analyzed the dataset collected from an ongoing long-term study \cite{agarwal_protectmyprivacy_2013} containing the app usage of a large number of iOS devices worldwide ($n = 166,006$). This dataset consists of fine grained app session length data (e.g. how long users used each app) from two different form factors --iPad and iPhone, over a 4-year long period from August 2012 to October 2016. To the best of our knowledge, this is one of the first studies that analyzed mobile device usage pattern at the scale of years.

% Introduce analysis plan
We take a systematic approach towards assessing the longitudinal variability in usage pattern across several dimensions. First we analyze App usage at a population-level and at an individual user level. Our population-level analysis of app usage, studies the global demand and needs of real-world ``average'' users, such as \emph{which apps are popular among the entire population and how this popularity has changed over time}. For our individual user-level analysis, we cannot get such popularity measures, but instead we focus on the interaction between users and their devices. For example, we explore research questions such as \emph{what is the number of apps users interact with during the course of a week?} or \emph{does the amount of time a user spends on a certain kind of apps changes over time?} Second, we study the longitudinal variability of app  usage at two different categorization levels (individual apps \& the category the app is listed under). We analyze category-level usage since the current categorization of Apps is somewhat based on high-level functionality (e.g. Games, Productivity, Travel), allowing us to highlight overall trends and how they have changed over time. We also study the longitudinal patterns of a set of top individual apps, because it allows us to look into user's behavior in greater detail by observing their actions. For example, by looking at what kinds of apps each user keeps using over time, we can refrain from recommending new apps of similar functionalities.

Based on our longitudinal analysis of app usage data across the above  dimensions, we make the following contributions:  

\begin{itemize}
    \item \textbf{Usage patterns at population level (\S \ref{section:population-level})}:
    \begin{itemize}
        \item \textbf{Proportional app category usage over time (\S \ref{subsection:proportional_category_usage}):} We show the proportional usage across app categories varied significantly for the first half of the dataset until the end of 2014. One notable change over that first period of time is a sharp decrease of productivity apps, led by decreasing use of mail apps. After 2015, we show that proportional usage across app categories stabilizes with relatively minor shifts such as usage of Photo \& video apps increasing while usage of productivity apps simultaneously decreasing. This kind of insights, along with other observations we make out of our dataset, app developers can be advised to develop apps with more visual entertainment than focusing on developing productivity apps. %This implies that we can expect the overall demand on mobile apps entered a stable stage.
        
        %\item \textbf{Proportional app category usage over time (\S \ref{subsection:proportional_category_usage}):} \done We show that there was a significant, longitudinal shift of general users' needs on mobile devices during the first half of our data collection period until the end of 2014. From 2015, we find they became stable with minor shifts such as a gradual increase in Photo \& video apps taking the proportional usage of productivity apps. This implies that we can expect the overall demand on mobile apps entered a stable stage.
        \item \textbf{App popularity (\S \ref{subsection:app_popularity}):} Even though the proportional usage across app categories remained stable after 2015, the proportional usage at the granularity of individual apps changed significantly over time. Given a list of top $N$ apps, we also find that roughly between $10\% \sim 20\%$ of those $N$ apps stay in that top $N$ list during the entire data collection period (Aug 2012 - Sept 2016), showing that these apps remain continuously popular. We also took a deeper look at which kinds of apps remain on the top apps for longer period of time, proving useful guidance to app developers on identifying what kinds of apps are less likely to be substituted by another new apps with similar functionality. For example, based on our analysis, Game apps typically had shorter life-time, failing to stay in the top list for an extended period of time by getting substituted by other Game apps quickly.
                
        %\item \textbf{App popularity (\S \ref{subsection:app_popularity}):} \done Even though the general needs on mobile apps remained stable after 2015, the proportional usage in the granularity of individual apps has gone through a significant change over time. Between $10\% \sim 20\%$ of the apps on the various monthly top $N$ app lists, appear on each of those lists for our entire data collection period (Aug 2012 - Sept 2016), showing that these apps are very popular. The various top $N$ app lists across consecutive months have a high level of similarity (i.e, average jaccard similarity of 0.64 on iPhone, 0.52 on iPhone month-by-month for the monthly list of the top-3000 apps), implying the list of top apps remain quite stable over time.
    \end{itemize}
    \item \textbf{Usage patterns at an individual user level (\S \ref{section:individual-level})}:
    \begin{itemize}
        \item \textbf{Individual variability in proportional usage over time (\S \ref{subsection:individual_usage_variance}):} We find that app usage pattern of each individual is highly variable over time, with regards to their proportional usage of apps in various categories. We find that each individual's usage pattern changes more dynamically than the global usage pattern changes,  by using simple, relative metrics to compare longitudinal variability of a typical, representative user and actual users in the dataset. %the standard deviation to mean ratio of month-by-month variation of 87.7\% and 89.9\% on iPhone and iPad, respectively, averaged across app categories. \YA{should this ratio not be a range, since you will have a number per month pair? Or maybe I don't understand what's being said here.}
        This individual variability may complicate building an effective app recommendation engine since frequent data collection of user app usage behavior to build accurate profiles is warranted. %is necessary to cope with temporal changes of its target users.
        \item \textbf{Working set (\S \ref{subsection:working_set}):} We find that the working set size of individual users is quite small. More than 90\% of the iPhone users in our dataset use between 14 to 18 different apps during the course of a  week (weekly working set size). In addition, we find the working set size for iPads is much smaller around 5 to 7 apps per week. For mobile platform designers and builders, these numbers provide a reference for deciding how many apps a mobile device should include in its default dashboard for better user experience. %In addition,  \YA{can say something more about why this information is useful?}  
    \end{itemize}
\end{itemize}
\section{Related Work}

Given the growth of mobile devices over the past decade, there have been numerous research efforts looking at all aspects of usage and app ecosystems surrounding them. Some of work attempts to leverage usage pattern for specific system objectives (e.g. improving network performance, optimizing energy use, providing better recommendations). The second line of work focuses on analysis of app usage patterns to answer fundamental user-app behavior questions. %To better explain the datasets and their findings, they often use various data analysis techniques such as clustering.

\subsection{Systems Using App Usage Data} 
Falaki et al. in their early work collected and studied app usage, as well as network and energy consumption data from 255 users, to show that usage patterns were diverse and could be leveraged for better predicting battery drain \cite{falaki_diversity_2010}. Yan et. al combined prior app usage with location and time of the day to predict future launches and pre-load apps for responsiveness \cite{yan_fast_2012}. Moodscope similarly used app usage with information such as phone calls, SMS, etc to predict a smartphone user's mood \cite{likamwa_moodscope:_2013}. CARAT uses crowd-sourced app usage data including CPU stats to alert users to ``hogs'', apps having unexpected energy use, an important issue on resource constrained devices, allowing users to take actions such as not using some apps or changing their device configuration \cite{oliner_sensys13_carat}. Several researchers have also built systems that use some combination of historical app usage \cite{shi_kdd2012_getjar}, contextual information from smartphone sensors such as location and social behavior, to predicting future app usage \cite{xu_preference_2013} along with recommending other applications that may be relevant \cite{shi_kdd2012_getjar, yan_mobisys11_appjoy}. In the context of mobile privacy, Liu et al. combine app usage data and privacy decisions of 72 smartphone users in a system providing a privacy ``nudge'' to users with similar profiles as others  \cite{liu_follow_2016}. 

\subsection{Studying App Usage Behavior}

More directly related to our work, prior research has studied app usage on mobile devices from a behavioral perspective. Li et al. for example  \cite{li_characterizing_2015} studied how users find, install, and remove apps, as well as the diversity in their network usage based on data from a Chinese app store and network usage traces for almost a million users. Xu et al. \cite{xu_identifying_2011} similarly use network traffic measurements from a US cellular provider, to study app usage for over 600,000 smartphone users. They show spatio-temporal usage correlations of apps, diurnal patterns of some apps, and usage correlation across apps. While their dataset was large, it was only collected for a week. At a smaller scale, Bohmer et al. studied 4,000 Android users' app usage over a 3-month period. They observed that smartphone users interacted with their device for an average of 59.23 minutes per day, with the mean app session lasting for 71.56 seconds \cite{bohmer_falling_2011}. However, they did not study changes in app usage behavior over time. On a smaller scale, Shin et. al study 48 participants over 25 days to detect abnormal app usage from a mental health standpoint \cite{shin_automatically_2013}. Eagle et al. demonstrated the ability to infer a wide variety of factors, such as relationships, socially significant locations, and organizational rhythms from 100 mobile phones over a period of 9 months \cite{eagle_reality_2006}. While this dataset is certainly long enough to examine longitudinal factors, such analysis was not the focus of Eagle et al's work. Do et al. also performed a small scale (111 participants) study based on 8-months of phone application usage to model and predict participant phone behavior \cite{do_by_2010}. More recently, the work by Hintze et. al. \cite{Hintze:2017} showed that location context (office, home, meaningful and elsewhere) as well as temporal context have a significant correlation with the usage pattern. Interestingly, since they used a dataset \cite{Wagner2014} with 4.5 years of data collection, they also found an evidence of longitudinal factors on usage pattern. However, they did not perform a detailed analysis on the longitudinal factors. Importantly, while they consider in their analysis that user behavior may change over time, and attempt to account for these changes, they did not perform in-depth analysis to quantify or examine to what extent this influences the usage behavior.

Researchers have also examined classifying users based on their app usage. Banovic et al. \cite{banovic_proactivetasks:_2014} observed the interaction of 27 users with the email app to classify them into four distinct types. Jones et al. \cite{jones_revisitation_2015} identified three groups of users based on a 165 participant, 3-month long study analyzing how users revisited the same apps. While these studies into app usage are interesting, the small scale and shorter study duration is a limitation given the sheer scale of mobile app usage. More recently, a study by Zhao et al. \cite{zhao_discovering_2016} studied a much larger dataset (n$\simeq$106,000) of Android users over a 1-month period to identify types of users. They identified 382 distinct user groups with notably distinct behaviors. While valuable and the basis for some of our work, their study had a relatively coarse-grained notion of usage, recording only the last ten apps used at the end of every hour, with no notion of how long those applications were used, or even launch frequency in that hour long period. Additionally, their relatively short data collection period prevented any longitudinal analysis of how users  change over time \cite{zhao_discovering_2016}. In this paper, we show that these longitudinal factors are key towards understanding app usage behavior over time.

To summarize, while there have been several studies of app usage, few have had large sets of diverse users, and among those with many subjects, few have collected data over a period long enough to examine longitudinal behavior changes. % Additionally, the recognized importance of such longitudinal app usage data can be seen in ongoing attempts to create longitudinal datasets. As an example, the 'mobile data challenge', an ongoing data collection and sharing effort by Nokia, specifically looks to create a longitudinal phone usage dataset that many researchers can use \cite{laurila_mobile_2012}. \YA{but, ... did they not release it? Its Nokia so may be irrelevant since they don't have an OS like Android or iOS?}
\section{Dataset}

In this section, we briefly introduce our dataset, how it was acquired, and describe the various pre-processing steps we took to protect participants' privacy while minimizing potential biases.

\subsection{Overview of Dataset}
\label{subsection:dataset}

To study longitudinal differences in app usage we use the dataset extracted from an ongoing long-term research study on mobile privacy \cite{agarwal_protectmyprivacy_2013, AndroidPMP}. These longer term research studies required low-level access to the smartphone OSes, which was not possible on vanilla unmodified devices. As a result, the ProtectMyPrivacy (PmP) App was developed for users with jailbroken iOS devices \cite{agarwal_protectmyprivacy_2013} and was made available in the Cydia app store, popular with tens of millions of iOS users at its peak. Our dataset also thus only includes usage records from jailbroken iOS devices. We acknowledge that this may lead to potential biases, and we discuss the implications later in this section (Section \ref{subsection:dataset_limitation}).

\textbf{App Session Record:} An \emph{app session} is the time period when the user puts a specific app in the foreground for interacting with it. In other words, an app session starts when a user clicks the app icon and ends when the user exits back to the home screen or pulls out another app on the foreground. In our dataset, each app session contains: \emph{(i)} the name of the launched app, \emph{(ii)} the app session start time in UTC, \emph{(iii)} the timezone information of the app session (i.e. time offset from system clock), \emph{(iv)} the date the app was launched and \emph{(v)} the duration for which it was in the foreground. In total, we had $1,369,862,864$ app session records recorded between August 2012 and October 2016. % \YA{from XX date to YY date, mention Month and years}

\textbf{Participants:}  Users in our dataset are real-world iOS users who voluntarily found and installed the ProtectMyPrivacy app. Under the approval from our institution's institutional review board (IRB), data was collected from only the users who explicitly consented inside the app.

\textbf{Population:}  Our dataset originally contains $244,587$ devices, but we filtered out devices that did not send any session data or only contained invalid session data. This pre-processing resulted in $177,165$ devices. Among those devices, there were $139,935$ (79.00\%) iPhones, $22,053$ (12.45\%) iPads, and $15,159$ (8.56\%) iPod-touch. In this paper, we focused on analyzing the usage data from the $166,006$ iPhones and iPads. 

 Note that no demographic details of any form, such as age, gender or nationality, were included in the dataset. Instead, to illustrate the geographical diversity of the users in this dataset, we extracted system timezone information included with the usage records, to show where the users are based at a coarse granularity. The numbers of devices categorized by timezone/location information in continental granularity are shown in Table \ref{tab:timezone_info}.
%We report the composition   (YA: this was an incomplete sentence, and hence was removed). 

\begin{table}
	\centering
	\parbox{0.48\textwidth}{
		\caption{\label{tab:timezone_info} Distribution of devices across location/timezones at the continental granularity. Since users may travel globally, or even fake their location, we performed equal-weighted sum for duplicate timezones recorded for a single device, in a way that each weight becomes one over the number of distinct duplicates. In ``Others'' category, we include Pacific, Atlantic, Antarctica times and other miscellaneous timezone/region values from which we could not extract location information (e.g. UTC-\#, GMT, etc). As can be observed, our dataset comes from users across the world, with close to half of our users coming from Asia.}
		\small
		\begin{tabular}{cc|cc}
			\toprule
			Timezone & Percentage & Timezone & Percentage \\
			\midrule
			Asia      & 45.92\% & Africa    & 2.13\% \\
			America   & 27.03\% & Oceania   & 1.68\% \\
			Europe    & 21.87\% & Others    & 1.37 \% \\
			\bottomrule
		\end{tabular}
		\vspace{0.6cm}
	}
	\hfill
	\parbox{0.48\textwidth}{
		\small
		\centering
		\caption{\label{table:app_labeling} Overview of App Labeling Process. \textsuperscript{1}Total number of apps in our dataset. \textsuperscript{2}Filtered out apps that had only invalid app sessions or no session data collected. \textsuperscript{3}Extracted app category information from the metadata recorded on each device at install time. \textsuperscript{4}Most appeared to be  development/test apps or apps with fewer than 4 users.}
		\begin{tabular}{clcc}
			\toprule
			Priority & Source & \# Labeled & \# Remaining\\
			\midrule
			- & Entire Dataset\textsuperscript{1}   & 0       & 481,073 \\
			- & Filter-out\textsuperscript{2}       & 4,088   & 476,985 \\
			\midrule
			1 & Manual                              & 3,000   & 473,985 \\
			2 & Apple AppStore                      & 232,372 & 241,613 \\
			3 & Cydia AppStore                      & 1,673   & 239,940 \\
			4 & Meta-data\textsuperscript{3}        & 188,192 & 51,748  \\
			5 & Others\textsuperscript{4}           & 51,748  & 0       \\
			\bottomrule
		\end{tabular}
	}
\end{table}

%\begin{table}
%\centering
%\begin{tabular}{cc|cc}
%\toprule
%Timezone & Percentage & Timezone & Percentage \\
%\midrule
%Asia      & 45.92\% & Africa    & 2.13\% \\
%America   & 27.03\% & Oceania   & 1.68\% \\
%Europe    & 21.87\% & Others    & 1.37 \% \\
%\bottomrule
%\caption{\label{tab:timezone_info} Distribution of devices across location/timezones at the continental granularity. Since users may travel globally, or even fake their location, we performed equal-weighted sum for duplicate timezones recorded for a single device, in a way that each weight becomes one over the number of distinct duplicates. In ``Others'' category, we had Pacific, Atlantic, Antarctica times and other miscellaneous timezone/region values from which we could not extract locational information (e.g. UTC-\#, GMT, etc).}
%\end{tabular}
%\end{table}

\subsection{Preprocessing}
\label{subsection:preprocessing}

Given the scale of our dataset, including data from a variety of jail-broken OS versions and different hardware devices, it is inevitable that some of these records constitute outliers which may not indicate actual app usage. To minimize the biasing effect of these outliers, we carefully designed a preprocessing procedure to filter out outlying app session data and robustly track each user across time and devices. % \YA{why is this referring to the same 3.2 section? maybe you mean 3.2.1?}

In addition, since our dataset contains a large number of distinct apps not only from Apple App Store but also from other sources including AppStores for jailbroken apps (Cydia AppStore\cite{Cydia}), we also carefully designed an app-labeling procedure detailed in Section \ref{subsection:app_labeling} for app-category level analysis.

\subsubsection{Session Filtering}

To prevent outlying app sessions from skewing our analysis, we removed all outliers (e.g. those sessions in the range of top 0.15\% and bottom 0.15\% of session lengths) from our analysis. In other words, we only considered the sessions that fall into the range of the center 99.7\%, leading us to only consider app usage records between 0.1959 seconds and 33189.9 seconds ($\sim 9.2$ hours) in length.
%\YA{we need some more justification why we chose these specific numbers?}. \YA{for example: 
We empirically chose the lower end cuttoff (0.15\% lowest session length) based on the assumption that the apps that are launched for less than 200ms are not really foreground apps and more likely a random keypress. We then chose a similar higher end cutoff (0.15\% highest session lengths), noting that apps that are in the foreground for 10 hours or more are quite likely unattended demo apps without any user interactions.

\subsubsection{Tracking Users Across Devices}
\label{subsubsection:device_filtering}

Since we collected usage logs for an extended period of time, the dataset not only reflects each individual user's app usage pattern change but also reflects each device's life cycle progression. For example, users can get a new device and potentially restore their state from an iCloud/iTunes backup. Moreover, it is also possible that users get did off their used device. These progression in each device's life cycle can potentially skew our data, because the dataset would include both the previous and new devices at the same time, regarded as two separate users in the dataset. To prevent these devices from skewing our dataset, we additionally and conservatively filtered devices that may not indicate the actual usage behavior.

Precisely tracking across multiple devices and their life cycles requires using personally identifiable information such as Apple ID. However, as noted in Section \ref{subsection:dataset}, the dataset does not collect any personally identifiable information (PII) from users. To track users in a way that does not require PII, we used a global unique identifier(GUID) generated \textit{locally on device} at install time in the app preference configuration. We call this GUID as the \emph{Install ID}. This install ID is automatically transferred to the new device when a user chooses to restore their own configuration or data from the previous one. We acknowledge that we cannot precisely track users across if they choose not to restore data, but we believe this procedure can alleviate the biasing effect of device life cycles.

In our dataset, we treated two devices with different hash values of device UUID as distinct devices. By default, we regard each distinct device is used by a unique owner (user) of that device. However, if two devices are of the same form factor and with the same install ID, we assumed the device is used by single user. When we analyze the data generated by those two devices, we only regarded the data from one of the devices that appeared later in time. Additionally, we also had cases where more than one install IDs were associated with single device UUID. We regarded this as the same user removing and later re-installing our PmP app on their device.

In this way, we identified 72,572 distinct users across 77,555 devices and 86,477 install IDs. We note that the overall population seems smaller now than the original 166,006 devices we started with, since the PmP app started to record install IDs in the middle of data collection period in January 2014. As a consequence, we lost a number of devices that stopped using the PmP app before this feature was introduced.

\subsubsection{Identifying Active Devices/Users:}
 When we were performing the longitudinal analysis, we found a number of devices showing very little activity to begin with. As those devices skewed our usage pattern substantially, we additionally filtered out devices without at least one app usage every other day given the time frame. For example,  if we were to filter out devices that do not seem to be active for a month, we checked whether the number of days a device had interactions exceeds the half of the days in that month. % \YA{Feb only or the year? readers may not know what month to month analysis is, and what Feb is wrt that? -- could we remove and add that in that analysis section?} 

\subsection{App Labeling}
\label{subsection:app_labeling}

%\begin{table}
%\centering
%\caption{\label{table:app_labeling} Overview of App Labeling Process. \textsuperscript{1}Total number of apps in our dataset. \textsuperscript{2}Filtered out apps that had only invalid app sessions or no session data collected. \textsuperscript{3}Extracted app category information from the metadata recorded on each device at install time. \textsuperscript{4}Most appeared to be  self-developed apps or apps with fewer than 4 users.}
%\small
%\begin{tabular}{clcc}
%\toprule
%Priority & Source & \# Labeled & \# Remaining\\
%\midrule
%- & Entire Dataset\textsuperscript{1}   & 0       & 481,073 \\
%- & Filter-out\textsuperscript{2}       & 4,088   & 476,985 \\
%\midrule
%1 & Manual                              & 3,000   & 473,985 \\
%2 & Apple AppStore                      & 232,372 & 241,613 \\
%3 & Cydia AppStore                      & 1,673   & 239,940 \\
%4 & Meta-data\textsuperscript{3}        & 188,192 & 51,748  \\
%5 & Others\textsuperscript{4}           & 51,748  & 0       \\
%\bottomrule
%\end{tabular}
%\end{table}

Since we have a large number of unique apps ($n=481,073$) in our dataset, we categorized each app into 25 different app categories to obtain higher-level insights on usage patterns. We accessed and used the same app categories defined by the iOS App Store at the end of 2016, to coincide with the end of our dataset, to cope with changes in Apple's categorization over time. Table \ref{table:app_category} provides the categories for reference. 

Given the large number of apps, we chose to utilize multiple sources of app labels instead of manually labeling them. We considered four different sources of app category labels: manual labeling, Apple AppStore, Cydia AppStore (an app store for jailbroken iOS devices) and meta-data residing with the app binary in each device. Because an app can have multiple labels from different sources, we carefully set up a priority based on the credibility and reliability of each source. The priority order and the number of apps labeled using each source are shown in Table \ref{table:app_labeling}. % \YA{meta point: In the future, we may just want to remove all the Cydia/JB apps from the analysis? that might be what is confusing reviewers too.}  

To ensure the correctness of app categorization, we manually labeled top 3000 most used apps in our dataset. This is based on the observation that app usage time is extremely skewed to a relatively small set of apps \cite{AndroidPMP}. As such, the top 3000 apps we manually labeled contribute about 91.34\% of the total usage time in our dataset (refer to Section \ref{subsection:app_popularity} for further details). For apps that did not appear any of the label sources mentioned earlier, we put them into a separate category we named \emph{Others}. These `other' apps contributed only 0.46 \% of app usage in the dataset.

\begin{table*}[t]
% Query: SELECT genre_id, COUNT(*) FROM app GROUP BY genre_id
\centering
\caption{\label{table:app_category}  Provides labels for Apple's App Store categories, as well as the popularity of the category (Number of apps used in that category) as well as example apps to give an understanding of the type of apps in each category.}
\scriptsize
\begin{tabular}{c c p{4.5cm} | c c p{4.5cm}} 
\toprule
Category & \# App & Example Apps & Category & \# App & Example Apps\\
\midrule
Business & 16,121 & Voxer, UberDriver, FedEx  &  Health\&Fitness & 11,145 & Daily Yoga, Fitocracy, Fitbit\\ 
Weather & 2,812 & DarkSky, Yahoo Weather, The Weather Channel & Games & 103,730 & Angry Birds, Infinity Blade, Sudoku \\
Utilities & 32,643 & Settings, Alarm Clock, Speed Test & Finance & 12,390 & Virtual Wallet, PNC Mobile, Discover Mobile \\
Travel & 16,492 & Expedia, Southwest, Trip Adviser & Entertainment & 36,428 & Netflix, HBO GO, Amazon Prime Video \\ 
Sports          & 9,351 & ESPN Sports, Yahoo Sports, Fox Sports & Education & 30,723 & iTunes U, Stack the States, Schoology \\
Social Networks & 16,376 & Facebook, Tumblr, OkCupid & Books & 13,850 & Kindle, iBooks, GoodReads \\
Reference       & 11,608 & Wolfram Alpha, Wikipedia, Dictionary.com & Medical & 6791 & Epocrates, UpToDate, Stress Check \\
Productivity    & 15,958 & Workflow, HabitList, SuperNotes & Newsstand & 82 & Marie Claire, Forbes Magazine, LA Times \\
Photo\&Video    & 17,089 & Camera, YouTube, PhotoVault & Catalogs & 1,918 & Classifieds, Tattoo Designs!, Perfumes \\
News            & 12,667 & Apple News, Google, TechCrunch& Food\&Drink& 5,925 & Starbucks, Wendy's, How To Cook Everything \\
Navigation      & 7,605 & Apple Maps, Google Maps, Waze & Shopping & 2,498 & Black Friday, SuperSaver, Woot \\
Music           & 15,747 & Spotify, Pandora, Google Music & Other & 52,476 & VUZIQ,iPhoneus, 'F--- You'\\
Lifestyle       & 28,648 & Tinder, Catholic Calendar, Reader's Digest & & &  \\
\bottomrule
\end{tabular}
\centering
\end{table*}

% SQL Queries
% Query: SELECT genre_id, COUNT(*) FROM app GROUP BY genre_id
% APPLE APPS SELECT COUNT(*) FROM app_statistics JOIN app ON app.id = app_statistics.app_id WHERE rank > 3000 AND JSON_CONTAINS_PATH(json, 'one', '$.genres');
% SQL - SELECT COUNT(*) FROM app_statistics JOIN app ON app.id = app_statistics.app_id WHERE rank > 3000 AND apple_genre_id IS NULL AND app_statistics.cydia_genre_id IS NULL AND (app.meta_genre_id IS NOT NULL OR (app.our_genre_id IS NOT NULL AND app.our_track_id IS NOT NULL))
% SELECT COUNT(*) FROM app_statistics JOIN app ON app.id = app_statistics.app_id WHERE rank > 3000 AND (NOT JSON_CONTAINS_PATH(json, 'one', '$.genres') or json IS NULL) AND cydia_genre_id IS NULL AND (meta_genre_id IS NOT NULL OR our_genre_id IS NOT NULL)

\subsection{Potential Bias in our App Usage Dataset}
\label{subsection:dataset_limitation}

We recognize the fact that our dataset has been collected from users of jailbroken iOS devices and my not be a representative sample of the entire population of regular non-jailbroken users. Jailbroken users are for example likely more technical than the average smartphone user and since our dataset is from users of the app for privacy, which means that the users are more privacy conscious \cite{agarwal_protectmyprivacy_2013}. Note, that as was the case for the original study, there is unfortunately no way to collect the app usage data that is needed for this large scale longitudinal analysis, at the granularity we need, without having a jailbroken device or otherwise modified OS. Apple has never exposed APIs to get fine grained app usage information on iOS and based on past experience has actively even removed apps that may have found indirect ways.
%\YA{Facebook app debacle, Onavo protect}.
We sincerely believe however that despite this potential bias, the dataset we have collected and analyzed is still useful to make several important observations based on overall trends of app usage across a four year data collection period. First, the Apps that these users use are primarily regular App store apps that they download from the regular App store. Second, while these users are likely more technical they are still somewhat representative of regular users who use standard apps and explore apps, switching between them and are subject to the same external signals (e.g. a new popular App being released, the Pokemon GO phase, etc) that regular non-jailbroken users are subject to that affect app usage. %\YA{Dohyun: reworded this please check} 

\section{Population-level App Usage Over Time}
\label{section:population-level}

In this section, we showcase our longitudinal analysis on app usage pattern at the entire population level. We first discuss the longitudinal shift in proportional usage time aggregated by app category (Section \ref{subsection:proportional_category_usage}). Then, we analyze the individual app level usage (Section \ref{subsection:app_popularity}), studying whether a typical user in our dataset spends the same amount of time on each app every month (Section \ref{subsubsection:app_popularity_by_usage_time}) and whether there is any substantial changes in the top $N$ lists over time (Section \ref{subsubsection:app_popularity_by_rank}). The first app-level analysis considers actual percentage of time a typical user spent using each app on her device. We also analyze the app popularity in terms of top $N$ most-used list to better understand the temporal dynamics of app popularity.

\subsection{Proportional App Category Usage over Time}
\label{subsection:proportional_category_usage}

\begin{figure*}[t]
\centering
\includegraphics[width=1\textwidth]{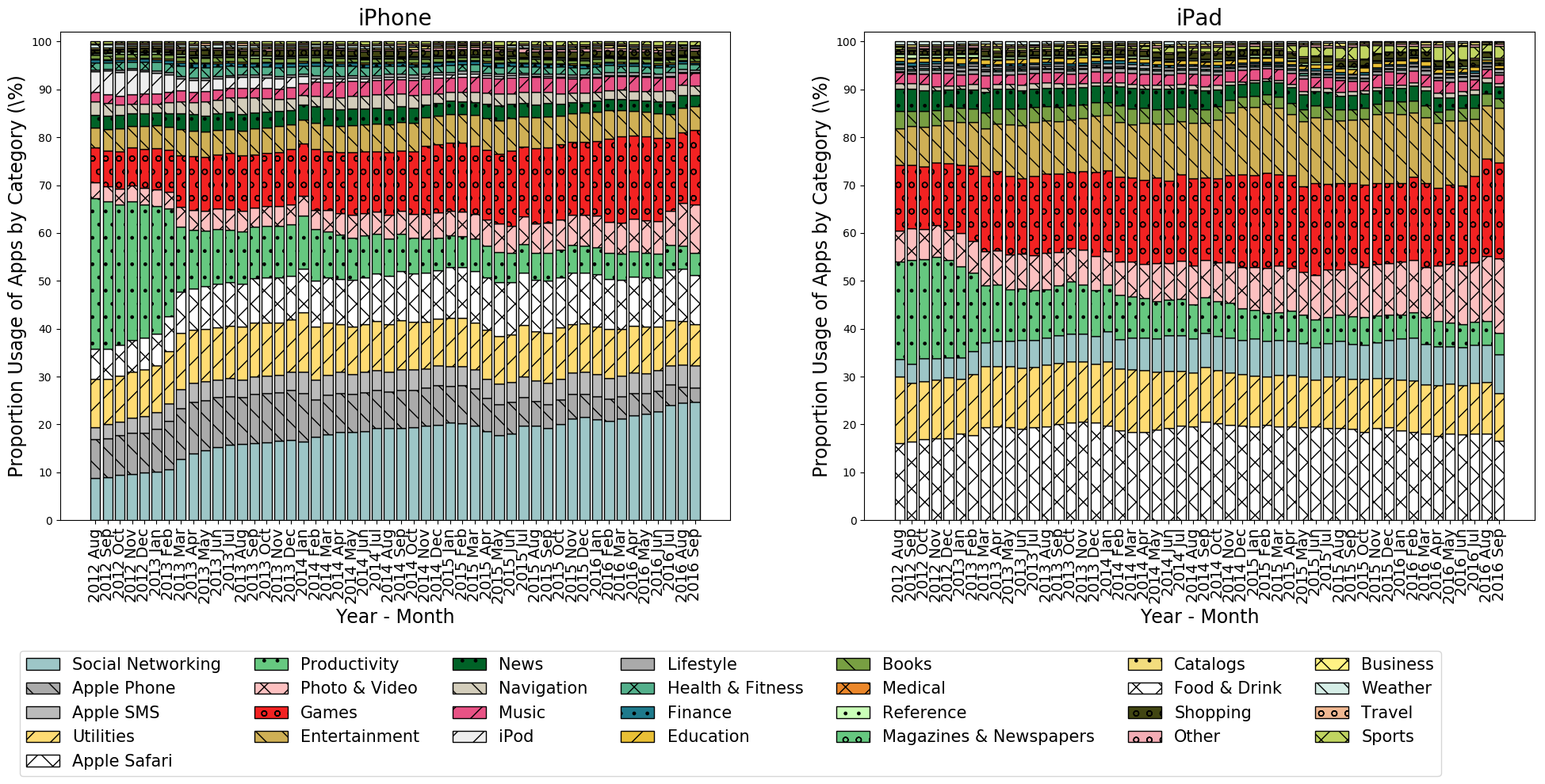}
\caption{\label{fig:overall_usage_per_genre_over_time} Proportional use of apps by category over Time. Use of various categories remains fairly stable after a relatively rapid shift that happened until the end of 2014. Notably, there was a significant difference between the two form factors. Social Networking apps was the top most used app on iPhone, whereas it was the 6-th most used app on iPad. Instead, Utilities, Games, and Entertainment apps were used more on iPads. Interestingly, the relative use of productivity apps shrunk significantly, while Games, Photo\&Video and Entertainment were used more frequently, regardless of form factor. In addition, because some of the Apple native-apps (Phone, SMS, iPod, Apple Safari) contributes a significant share of app usage they are categorized to, we depicted those apps separately.}
\end{figure*}

We first report on the proportional usage averaged across all months to show that the usage patterns for the iPhone and iPad were significantly different. For example, Social Networking apps contributed 29.1\% of the proportional usage on the iPhone, as compared to only 7.2\% of overall usage on the iPad. This is not just because iPhone has Phone and SMS apps, because Phone and SMS app itself contributes 7.3\% and 3.9\%, leaving 17.7\% of other social networking app usage. This is still higher than the whole Social Networking app usage percentage in iPads. This resulted in Social Networking being the top most category on iPhone and the sixth most used app category on the iPad. Instead, four other app categories (i.e. Utility, Games, Entertainment and Photo\&Video) showed higher proportional usage on iPad. More specifically, iPad users spent time on Utility, Games, Entertainment and Photo\&Video apps for 30.5\%, 17.0\%, 11.8\% and 8.8\% of their time while iPhone users spent 19.2\%, 12.9\%, 5.6\% and 5.1\% on those apps, respectively. For Utility apps other than Safari, there was no significant difference in Utility app use (11.6\% on iPads and 10.5\% on iPhones). However, Safari was intensively used on iPads (18.9\%), making huge difference in proportional usage of Utilities apps. On iPhone, users spend 9.5\% of the time using Safari. %\YA{Just curious: is the utilities app on the iPAD include the browser? might want to say it in the app level graph? this is in fact what the reviwer-1 commented below and we should bring that out.} 

The average usage pattern was impacted not only by the form factor, but also by the long-term temporal context. Figure \ref{fig:overall_usage_per_genre_over_time} shows the proportional time time an average user in our dataset used apps from each category, for both form factors. The usage pattern was split on a monthly basis over four years (August 2012 to October 2016). We found that the usage of Productivity apps (lightgreen with dot shade in Figure \ref{fig:overall_usage_per_genre_over_time}) had been notably trending down for both of the form factors. Productivity apps contributed 31.5\% and 20.4\% of overall app usage in August 2012 on iPhone and iPad respectively, but ended with 4.53\% and 4.45\% each in September 2016. In contrast, users spent more time using apps of several other categories than they had spent on them four years before. The typical example was Photo\&Video apps whose usage grew from 3.2\% and 6.5\% in August 2012 to 10.2\% and 15.6\% in September 2016, respectively.

The longitudinal change in usage for Games and Entertainment apps is interesting since both showed positive growth, but the rate of growth differed on the two form factors. On the iPhone, the usage of Game Apps grew from 7.3\% in August 2012 to 15.4\% in September 2016. iPad usage increase was more modest: 13.6\% in August 2012 and 20.0\% in September 2016. In contrast, on iPhone, Entertainment apps usage increased from 4.1\% in August 2012 to 4.9\% in October 2016, while on the iPad the increase during the same period was more significant rising from 7.7\% to 11,3\%. Our data shows that the rate of usage pattern change could vary significantly between form factors, even though the trends may be similar.

Overall, we find that there is a longitudinal changes in app usage pattern in app category-level. Notably, the change was more faster until the end of 2014 and the change slowed down starting from January 2015. % \YA{I don't get the first half part? maybe it needs to be reworded here and also in the intro?}
We also observe that users spent more time on entertainment (using Games, Photo \& Video and Entertainment apps) than before, while there was a significant decline in Utility app usage. This trend was similar on both form factors, whereas they have different baseline per-category usage patterns.
% \Reviewer{(1) I am not sure	whether the finding that app use has shifted over time is an indicator for shifted user needs. I'd rather believe that app use changes because apps are not yet old and new apps can still rise to a dominant market share. (2) There was always a need for entertainment, but only recently, YouTube has become sufficiently interesting -- hence the rise in YouTube consumption. (3) the 18.5 \% in the	Utilities categories does not mean that Utility apps are popular, it just means that people browse a lot with Safari (assuming that Safari falls under Utilities).}

\subsection{App Popularity}
\label{subsection:app_popularity}

\begin{figure*}[t]
	\centering
	\includegraphics[width=1\textwidth]{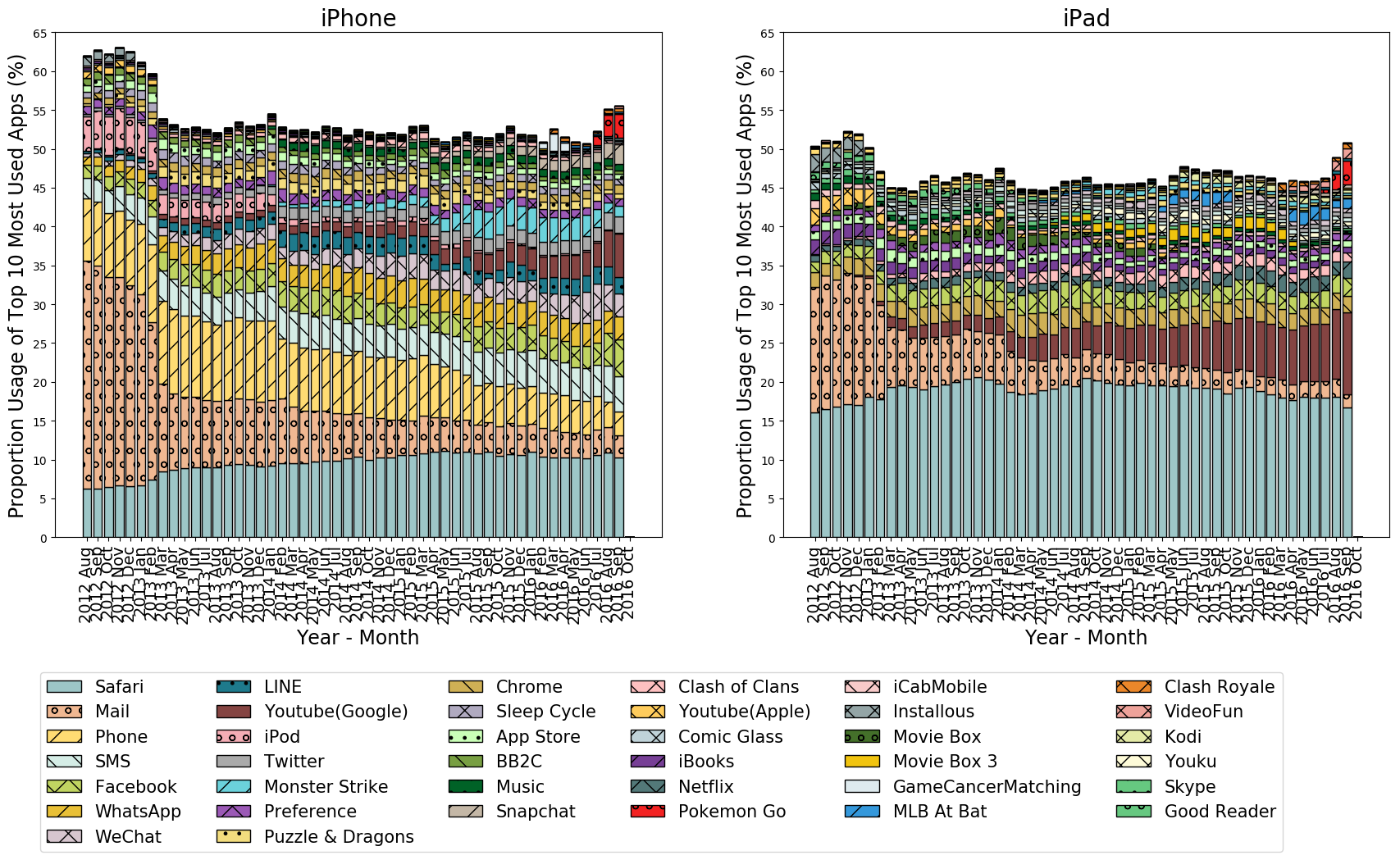}
	\caption{\label{fig:overall_usage_per_app_over_time} Proportional usage of apps over time. We find that app usage is extremely skewed so that only a small subset of apps ($n=38$) account for a significant proportion of total app usage. There was a significant change in app usage over time. For example, there was a consistent decline in Apple Mail app over time while a gradual growth of Youtube app.}
\end{figure*}

\subsubsection{App Popularity by Usage Time} \label{subsubsection:app_popularity_by_usage_time}

To analyze the longitudinal effects on app usage over time, we considered using the app usage time per month as a feature. However, we discarded this as being infeasible due to the sheer number of apps ($n=481,073$). Furthermore, we observed that app usage is highly skewed such that a significantly smaller subset of apps ($n=38$) account for a significant proportion of total app usage (46.8\% on the iPad and 52.6\% on the iPhone). This result indicates that users used a small set of apps extensively with a significant overlap between the working set of apps between users. It also indicates that iPhone users have more skewed usage of apps than iPad users. The process we followed to get to these popular apps is the following: (i) we calculate the top-10 most used apps on a monthly basis for both form factors; (ii) concatenate these monthly top-10 app lists, and (iii) removed duplicates so that each unique app occurs once. In doing so, we ensure that any app that was in the top-10 list at any point, for either form factor, is included.

We see a significant difference between iPad's usage and iPhone's usage. As expected, iPhone users spent significant proportion of their time on phone calls and using SMS. In contrast, these apps were not extensively used on iPads, since most iPads users tend not to have cellular modems.

Also, generally we observed more media-oriented usage on iPads than on iPhone. Other than Phone and SMS apps, we found more extensive use of social network apps (Facebook, WhatsApp, WeChat, Twitter, and LINE) on iPhones. Their usage is fairly stable over time, without any significant increase or decrease of usage. However, only Facebook app showed up on iPads as an extensively used app. This is because of the difference between Facebook and the other dominant Social Networking apps. Facebook is more media-oriented (capable of video/image sharing), while the other apps are more text-based (messaging). Also, we found more usage of video apps (e.g. Youtube, Netflix) and web browsers(e.g. Safari, Chrome) on iPads than on iPhone.

There were a few other interesting longitudinal changes over time. Across the two form factors, we observe a remarkable decline of Apple Mail app usage. %This indicates the overall decline of e-mail usage in general because there was no significant increase of its substitute apps such as Microsoft outlook for iOS and Google's Gmail client. \YA{is this really verified? was this replaced by IM and other apps?}
During the same period of time, iPhones and iPads both experienced an increase in usage of Youtube apps (brown unshaded from Figure \ref{fig:overall_usage_per_app_over_time}). In addition, we observe a significant usage of PokemonGO(red, shaded with large circles around right top corner of Figure \ref{fig:overall_usage_per_genre_over_time}) from July 2016 to September 2016. On iPhone, the app was the seventh(3.02\%) most used app in August 2016. The similar pattern was found on iPad: PokemonGo was the fourth (3.07\%) most used app on September 2016 on iPad.

In summary, even though we have seen stable app category-level usage pattern from the start of 2015, but our in-depth analysis on app-level usage shows a different picture of longitudinal shift.

\subsubsection{App Popularity in top $N$ lists}
\label{subsubsection:app_popularity_by_rank}

\begin{figure*}[t]
	\centering
	\hfill
	\begin{minipage}[p]{.48\textwidth}
	\includegraphics[width=\textwidth]{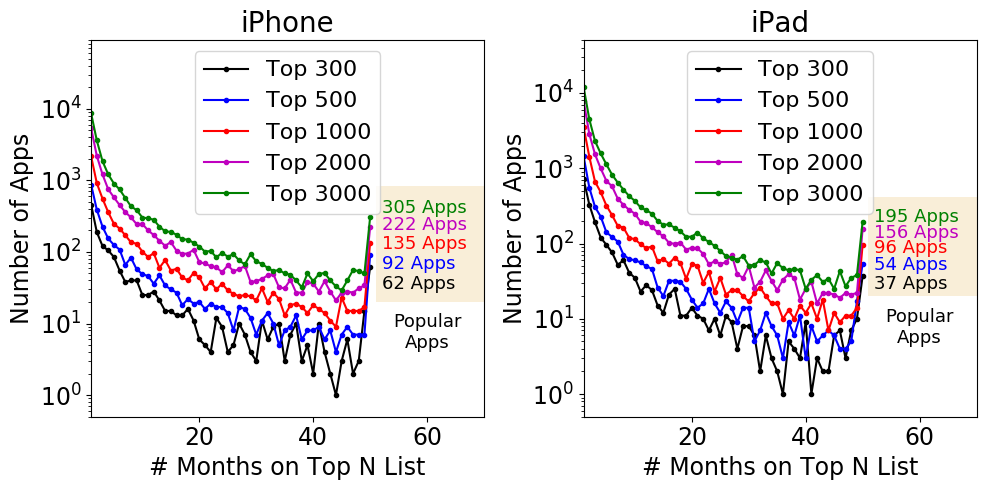}
	\caption{\label{figure:longitudinal_app_popularity} App Popularity over Time. On the list of top-N apps, we have about $10\% \sim 20\%$ among those N apps appearing on the top-N app list for the whole period of data collection. Those apps are noted as ``Popular Apps'' in the diagram. For example, for top 3000 apps, 305 iPhone apps and 195 iPads apps were on the list for the whole data collection period.}
	\end{minipage}
	\hfill
    \begin{minipage}[p]{.48\textwidth}
 		\includegraphics[width=\textwidth]{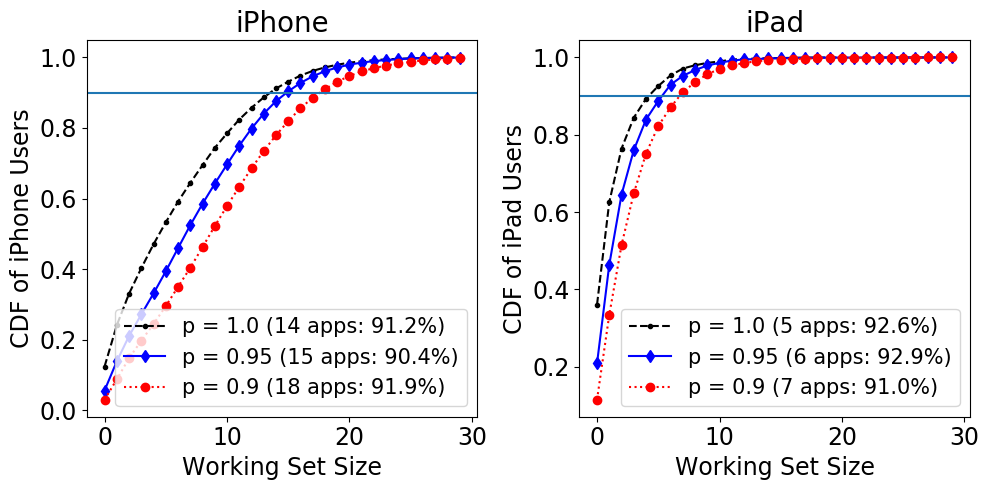}
		\caption{\label{figure:user_working_set} Size of User Working Set. $p$ is a probability of an app being used for $100p\%$ of the weeks at which a certain user is active. \emph{Working Set} is defined as the number of apps that appear on user's weekly launched app list more than $100p\%$ of weeks. We find that if we use $p=1.0$ where a working set is the number of apps that have been launched every single week, 91.2\% of the users had less than or equal to 14 apps.}
    \end{minipage}
	\hfill
\end{figure*}

 Next we analyze longitudinal patterns of app popularity, i.e. \emph{does app popularity change over time?}. The goal of this analysis is to see whether apps released later have a disadvantage in terms of gaining popularity as compared to older entrenched apps. In this section, we investigate how the set of most used apps change over time. 

\textbf{Longitudinally Popular Apps:}  We first extracted each month's top $n$ list of apps ($n=300, 500, 1000, 2000, 3000$) based on usage time. Then, we calculated the number of months each of these apps stayed in the respective top-n list throughout the timespan of our dataset. We find that a significant number of apps (10\% $\sim$ 20\% of top $n$ apps) remained in the top list for the entire date collection period (50 months). For example, in the set of the top $n=300$ apps for iPhone's and iPad's, we found that 62 iPhone apps (20.6\% of the top 300 apps) and 37 iPad apps (12.3\% of the top 300 apps) are consistently seen in each month's usage, for the entire period. We see a similar pattern for the top $500, 1000, 2000, 3000$ lists. We call these apps that are seen in the top list every single month as \emph{longitudinally popular apps} for the rest of this section.

 Figure \ref{figure:longitudinal_app_popularity} shows the number of apps (y-axis) that stayed on the various top-n App lists, for how many months (x-axis), across our data collection period of 50 months. The apps that were present on user's devices before our data collection started (August 2012) and remain at the end (Oct 2016) form a set of ``Popular Apps'', marked on the right side of the graph. As an example, 62 apps were in the top-300 popular Apps across the entire period. In contrast, we see a steady monotonic decrease in the set of apps in the various top-n lists, as the length of time when they remain popular increases from 1 month to 50 months. This is expected, as a fewer set of apps can remain popular over longer periods of time. 

\textbf{Deeper look at longitudinally popular apps:}  Next we performed a more detailed case study on the top $300$ App list to understand the characteristics of ``longitudinally popular apps'', i.e. apps that remain on the top-300 list for the entire 50 month period. \emph{We find that about one third of these apps are Apple native (pre-installed) apps, indicating their popularity on iOS.} Specifically, we see 16 apps on the iPhone (32.4\% of 62 Apps) and 12 apps on the iPad (25\% of 37 Apps) are developed by Apple. Second, we observe that about half of these longitudinally popular apps are Utility and Social Networking apps (51.6\% on iPhone and 40.5\% on iPad). However, these Utility and Social Networking Apps are not as popular when we consider the entire set of apps that have appeared on the top-300 app list \emph{at least once}. There were 2012 apps (iPhone) and 1556 apps (iPad) that have shown up on the list at least once over the years and 47.8\% and 44.0\% of these apps were apps categorized as a Game on iPhones and iPads, respectively. As such, Game apps stayed in the top-300 list, for 5.7 months(iPhone) and 4.5 months(iPad) on average, Social Networking Apps for 14.4 months (iPhone) and 14.5 months (iPad), and Utility Apps for 11.0 months (iPhone) and 11.7 months. This indicates that \emph{Social Networking and Utility apps dominate when we consider longitudinal popularity, while Game apps tend to have comparably shorter periods of popularity even though they show up on the top-300 lists in multiple months.}

\textbf{Monthly Change of Popular Apps:}  We observed earlier that about 10\% - 20\% of the apps appear in the top $N$ lists across our entire dataset (50 month). Next we study the month to month variation of these top-$N$ most used App lists, to understand how different, or similar, they are. Specifically, we calculate the Jaccard Similarity of each top $N$ list for consecutive months to see whether there is a significant change at the entire population level. In particular, we use the Jaccard similarity measure $\mathcal{J}$ which is defined as $\mathcal{J} = \frac{|\mathcal{A}_{n} \cap \mathcal{A}_{n+1}|}{|\mathcal{A}_{n} \cup \mathcal{A}_{n+1}|}$ where $\mathcal{A}_{n}$ is the top $N$ list of popular apps in the $n$-th month. If the sets of top $N$ apps in consecutive months are the same, the $\mathcal{J}$ value is $1.0$ (the higher $\mathcal{J}$, the more similar the sets are).

When considering just the top-10 apps, we see average Jaccard similarities of 0.85 (iPhone) and 0.86 (iPad) over the 50 month period, thereby showing high overlap across consecutive months. When looking at the list of top-1000 apps, we observe a Jaccard similarity value of 0.64 (iPhone) and 0.52 (iPad) over our data collection period. We did not find any meaningful longitudinal pattern such as increasing or decreasing Jaccard similarity over time.

\section{Individual-level App Usage Over Time}
\label{section:individual-level}

The above sections illustrate how our user population behaved as a whole, including how their app usage behavior changed over time. However, while potentially useful, it is important to consider the extent to which individuals varied over time. In Section \ref{subsection:individual_usage_variance}, we investigate whether each individual's usage pattern has changed over time. Then, we move on to working set analysis in Section \ref{subsection:working_set}.

\subsection{Individual Usage Variance in Monthly Usage}
\label{subsection:individual_usage_variance}

\begin{table*}[t]
	\centering
	\caption{\label{tab:SingleVarariance}  \textsuperscript{1}The standard deviation of each individual's usage over time per each category, averaged across all users. \textsuperscript{2} The average \% difference of the standard deviation in usage from mean usage over time (month to month) for each category, averaged across all users. \textsuperscript{3} The standard deviation of a typical user's usage. \textsuperscript{4} The typical user's \% difference of the standard deviation in usage from mean usage over time (month to month) for each category. High values highlighted in red and marked with $^a$. Low values highlighted in blue and marked with $^b$. Users had a high average variability across categories (87.7\%  average \% change on iPhones and 98.9\% average \% change on iPads). This varied with categories, as variability of categories in Blue$^b$ were much lower than those in Red$^a$. In general, the app categories with more usage had higher standard deviation and lower average change over time. Also, we observe that a typical user's variability over time is way lower than that of individual users in the dataset, indicating that individual's longitudinal variability extends beyond just the typical usage change. This high degree of longitudinal variability was observed across form factors.}
	\scriptsize
	\begin{tabular}{l|llll|llll}
		\toprule
		\multirow{3}{*}{Category} & \multicolumn{4}{c|}{iPhone} & \multicolumn{4}{c}{iPad} \\
		& \multicolumn{2}{c}{Average across Users} & \multicolumn{2}{c|}{Typical User} & \multicolumn{2}{c}{Average across Users} & \multicolumn{2}{c}{Typical User} \\
		& Avg. S.D.(\%)\textsuperscript{1} & Avg. Change(\%)\textsuperscript{2} & S.D.\textsuperscript{3}(\%) & Change\textsuperscript{4}(\%) &  Avg. S.D.(\%)\textsuperscript{1} & Avg. Change(\%)\textsuperscript{2} & S.D.\textsuperscript{3}(\%) & Change\textsuperscript{4}(\%) \\
		\midrule
		Business          & \hphantom{0000}0.33                          & \hphantom{0000}111.9                       & \hphantom{00}0.064                          & \hphantom{0000}19.5
    	& \hphantom{0000}0.45                          & \hphantom{0000}127.0                       & \hphantom{00}0.072                         & \hphantom{0000}18.3            \\
		Weather           & \hphantom{0000}0.27                          & \hphantom{0000}103.5                       & \hphantom{00}0.12                         & \hphantom{0000}34.9
		& \hphantom{0000}0.27                          & \hphantom{00000}92.0                       & \hphantom{00}0.16                          & \hphantom{0000}47,8            \\
		Utilities         & \hphantom{0000}\textcolor{red}{$6.56^a$}     & \hphantom{00000}\textcolor{blue}{$39.8^b$} & \hphantom{00}\textcolor{red}{$1.7^a$}     & \hphantom{0000}\textcolor{blue}{$16.5^b$}
		& \hphantom{000}\textcolor{red}{$10.2^a$}     & \hphantom{00000}46.2                       & \hphantom{00}\textcolor{red}{$1.08^a$}     & \hphantom{00000}\textcolor{blue}{$9.3^b$} \\
		Travel            & \hphantom{0000}0.29                          & \hphantom{0000}107.9                       & \hphantom{00}0.051                         & \hphantom{0000}\textcolor{blue}{$16.5^b$}
		& \hphantom{0000}0.31                          & \hphantom{0000}\textcolor{blue}{$120.5^b$} & \hphantom{00}0.052                         & \hphantom{0000}20.8            \\
		Sports            & \hphantom{0000}0.37                          & \hphantom{00000}65.2                       & \hphantom{00}0.16                          & \hphantom{0000}36.7
		& \hphantom{0000}1.08                          & \hphantom{00000}73.2                       & \hphantom{00}0.84                          & \hphantom{0000}\textcolor{red}{$77.8^a$}            \\
		Social Networking & \hphantom{0000}\textcolor{red}{$7.56^a$}     & \hphantom{00000}\textcolor{blue}{$40.2^b$} & \hphantom{00}\textcolor{red}{$4.23^a$}     & \hphantom{0000}\textcolor{blue}{$23.7^b$}
		& \hphantom{0000}4.19                          & \hphantom{0000}\textcolor{blue}{$109.3^b$} & \hphantom{00}1.28                          & \hphantom{0000}\textcolor{blue}{$19.4^b$}            \\
		Reference         & \hphantom{0000}0.26                          & \hphantom{0000}107.9                       & \hphantom{00}0.078                         & \hphantom{0000}25.9
		& \hphantom{0000}0.42                          & \hphantom{0000}117.2                       & \hphantom{00}0.10                         & \hphantom{0000}23.5            \\
		Productivity      & \hphantom{0000}\textcolor{red}{$4.16^a$}     & \hphantom{00000}73.7                       & \hphantom{00}\textcolor{red}{$7.51^a$}     & \hphantom{0000}\textcolor{red}{$69.2^a$}
		& \hphantom{0000}\textcolor{red}{$5.05^a$}     & \hphantom{0000}102.1                      & \hphantom{00}\textcolor{red}{$4.88^a$}     & \hphantom{0000}53.6            \\
		Photo\&Video      & \hphantom{0000}2.85                          & \hphantom{00000}65.8                       & \hphantom{00}1.46                          & \hphantom{0000}28.6
		& \hphantom{0000}5.70                          & \hphantom{00000}96.2                       & \hphantom{00}\textcolor{red}{$2.11^a$}     & \hphantom{0000}24.0            \\
		News              & \hphantom{0000}1.36                          & \hphantom{00000}84.0                       & \hphantom{00}0.44                          & \hphantom{0000}\textcolor{blue}{$15.2^b$}
		& \hphantom{0000}1.75                          & \hphantom{00000}98.0                       & \hphantom{00}0.62                          & \hphantom{0000}17.6            \\
		Navigation        & \hphantom{0000}1.69                          & \hphantom{0000}100.1                       & \hphantom{00}0.47                          & \hphantom{0000}\textcolor{blue}{$20.1^b$}
		& \hphantom{0000}0.85                          & \hphantom{0000}143.0                       & \hphantom{00}0.18                          & \hphantom{0000}21.9            \\
		Music             & \hphantom{0000}2.46                          & \hphantom{0000}100.1                      & \hphantom{00}0.52                         & \hphantom{0000}19.4
		& \hphantom{0000}2.18                          & \hphantom{0000}\textcolor{red}{$144.2^a$}  & \hphantom{00}0.23                          & \hphantom{0000}10.0            \\
		LifeStyle         & \hphantom{0000}0.48                          & \hphantom{0000}\textcolor{red}{$113.7^a$}  & \hphantom{00}0.47                          & \hphantom{0000}\textcolor{red}{$80.5^a$}
		& \hphantom{0000}0.70                          & \hphantom{0000}\textcolor{red}{$126.6^a$}  & \hphantom{00}0.12                          & \hphantom{0000}18.0            \\
		Health \& Fitness & \hphantom{0000}1.08                          & \hphantom{0000}116.8                       & \hphantom{00}0.36                          & \hphantom{0000}22.7
		& \hphantom{0000}0.30                          & \hphantom{00000}90.0                        & \hphantom{00}0.069                         & \hphantom{0000}27.9            \\
		Games             & \hphantom{0000}\textcolor{red}{$6.11^a$}     & \hphantom{00000}97.7                       & \hphantom{00}\textcolor{red}{$2.87^a$}     & \hphantom{0000}22.2
		& \hphantom{0000}\textcolor{red}{$8.47^a$}     & \hphantom{0000}103.0                       & \hphantom{00}1.72                          & \hphantom{0000}\textcolor{blue}{$10.1^b$} \\
		Finance           & \hphantom{0000}0.44                          & \hphantom{0000}107.9                       & \hphantom{00}0.12                         & \hphantom{0000}19.6
		& \hphantom{0000}0.37                          & \hphantom{00000}95.7                       & \hphantom{00}0.094                         & \hphantom{0000}23.1            \\
		Entertainment     & \hphantom{0000}2.97                          & \hphantom{0000}100.7                       & \hphantom{00}0.61                          & \hphantom{0000}11.0
		& \hphantom{0000}\textcolor{red}{$7.16^a$}     & \hphantom{0000}106.2                       & \hphantom{00}\textcolor{red}{$1.88^a$}     & \hphantom{0000}15.8            \\
		Education         & \hphantom{0000}0.28                          & \hphantom{0000}\textcolor{red}{$114.9^a$}  & \hphantom{00}0.066                         & \hphantom{0000}27.0
		& \hphantom{0000}0.90                          & \hphantom{0000}127.5                       & \hphantom{00}0.12                          & \hphantom{0000}\textcolor{blue}{$14.2^b$}            \\
		Books             & \hphantom{0000}0.69                          & \hphantom{0000}\textcolor{red}{$117.4^a$}  & \hphantom{00}0.14                          & \hphantom{0000}18.3
		& \hphantom{0000}2.38                          & \hphantom{0000}134.3                      & \hphantom{00}0.43                          & \hphantom{0000}15.7            \\
		Medical           & \hphantom{0000}\textcolor{blue}{$0.095^b$}   & \hphantom{00000}56.0                       & \hphantom{00}\textcolor{blue}{$0.029^b$}   & \hphantom{0000}36.5
		& \hphantom{0000}\textcolor{blue}{$0.097^b$}   & \hphantom{00000}\textcolor{red}{$54.0^a$}  & \hphantom{00}\textcolor{blue}{$0.043^b$}   & \hphantom{0000}\textcolor{red}{$45.9^a$}            \\
		Newsstand         & \hphantom{0000}\textcolor{blue}{$0.00023^b$} & \hphantom{000000}\textcolor{blue}{$1.82^b$} & \hphantom{00}\textcolor{blue}{$0.00023^b$} & \hphantom{000}\textcolor{red}{$119.4^a$}
		& \hphantom{0000}\textcolor{blue}{$0.0043^b$}  & \hphantom{000000}4.22                       & \hphantom{00}\textcolor{blue}{$0.0036^b$}  & \hphantom{000}\textcolor{red}{$113.3^a$}            \\
		Catalogs          & \hphantom{0000}\textcolor{blue}{$0.021^b$}   & \hphantom{00000}\textcolor{blue}{$45.1^b$} & \hphantom{00}\textcolor{blue}{$0.0046^b$}  & \hphantom{0000}29.4
		& \hphantom{0000}\textcolor{blue}{$0.031^b$}   & \hphantom{00000}\textcolor{blue}{$39.4^b$} & \hphantom{00}\textcolor{blue}{$0.0093^b$}   & \hphantom{0000}\textcolor{red}{$49.4^a$}            \\
		Food\&Drink       & \hphantom{0000}\textcolor{blue}{$0.13^b$}    & \hphantom{00000}94.8                       & \hphantom{00}\textcolor{blue}{$0.16^b$}   & \hphantom{000}103.2
		& \hphantom{0000}\textcolor{blue}{$0.16^b$}    & \hphantom{00000}\textcolor{blue}{$87.5^b$} & \hphantom{00}\textcolor{blue}{$0.031^b$}   & \hphantom{0000}25.0            \\
		Shopping          & \hphantom{0000}0.74                          & \hphantom{00000}93.7                       & \hphantom{00}0.25                          & \hphantom{0000}25.5
		& \hphantom{0000}0.82                          & \hphantom{00000}98.7                       & \hphantom{00}0.16                          & \hphantom{0000}16.9            \\
		Other             & \hphantom{0000}0.47                          & \hphantom{0000}\textcolor{red}{$132.2^a$}  & \hphantom{00}0.17                          & \hphantom{0000}37.8
		& \hphantom{0000}0.47                          & \hphantom{0000}\textcolor{red}{$137.1^a$}  & \hphantom{00}0.089                          & \hphantom{0000}27.2            \\
		Average           & \hphantom{0000}1.67                          & \hphantom{00000}87.7                       &  \hphantom{00}0.88                         & \hphantom{0000}35.2  
		& \hphantom{0000}2.18                          & \hphantom{00000}98.9                       & \hphantom{00}0.65 & \hphantom{0000}29.9 \\
		\bottomrule
		%\textcolor{red}{$^a$} \textcolor{blue}{$^b$} 
	\end{tabular}
	\centering
\end{table*}

To investigate each individual's usage pattern, we extracted the proportional time each user spent on each of the app categories every month. We then calculated the standard deviation and the average of these proportions per app category across time. We utilized two different metrics to assess longitudinal variability: the native standard deviation as an objective metric of the average variability of a device across time as well as the proportion of the standard deviation to the mean (Std-Dev/Mean*100). The latter metric, which we will call ``\% change from mean'' henceforth in this section, is an indicator of the relative scale of the variability to the overall baseline usage. For example, if ``\% change from mean`` of an app category usage is higher than that of other over time, it means that the usage pattern of that app category has more temporal variability than the other app category's usage. Importantly, we only consider data for those users with more than one month of data in our dataset, as a Std-Dev with 1 data point does not provide meaningful insight and could potentially skew our results.

We then average these two metrics across all of the users, to assess the average level of each individual device's variability across our dataset. Additionally, we represented the results with the standard deviation and \% change value of an average user for reference, which is already shown in a form of breakdowns in Figure \ref{fig:overall_usage_per_genre_over_time}.

The results of this analysis can be seen in Table \ref{tab:SingleVarariance}. Across all categories, users had an average standard deviation of 1.67\% and 2.18\%, with an average ratio between standard deviation and mean (Std-Dev/Mean *100) of 87.7\% and 98.9\% on iPhones and iPads, respectively. These numbers are roughly or more than twice the standard deviation and \% change from mean value of a typical user shown in Figure \ref{fig:overall_usage_per_genre_over_time}. This means that individual user's usage pattern changes more dramatically than it does at the population level. In other words, if each individual's longitudinal app usage pattern has changed over time at the same rate of the global change, standard deviation and change from mean values would have been on a similar level. This leads us to a deeper insight into each individual's usage pattern over time: if we draw a graph similar to Figure \ref{fig:overall_usage_per_genre_over_time} based on each individual's usage behavior, we would end up with a graph with larger longitudinal variation across time. 

Examining the results in Table \ref{tab:SingleVarariance} in further detail, we found that app categories with more usage (i.e. Social Networking, Utility, Productivity) have larger variance over time, while others with less usage such as Medical, Newsstand, Catalogs and Food\&Drink have smaller variance over time. This is expected because the standard deviation of a data reflects the scale of the original data points. However, despite their high level of standard deviation in its absolute value, Utility and Social Networking apps on iPhone have the lowest \% change from mean. This indicates their usage is relatively stable over time, although they still suffer from longitudinal variation of 40.2\%(Social Networking) and 39.8\%(Utility) from mean on iPhone. 

This pattern was observed both as an individual user and an average user. However, we could not find the similar pattern on iPads from each individual's usage; the \% change from mean of the top most used app categories on iPad--Utility, Games and Entertainment--were not the lowest ones, rather being the ones around the average. This implies that there are differences in the magnitude of longitudinal changes among different form factors and app categories.

\subsection{Working Set of Individual Users}
\label{subsection:working_set}

 As we previously found that the list of popular apps remain similar across the entire data collection period in Section \ref{subsubsection:app_popularity_by_rank}, we next study whether the cause for this is the relative stability of the set of apps individual users use over time. 

\textbf{Extracting Working Set Size from Users:}  We defined the \emph{working set of a user} for a certain time period as: \emph{the list of apps a user launches during that given period.} Since selecting a time period for this analysis is important, we chose to use \emph{a week} since it covers both diurnal and weekday/weekend variations, and is also not too long. To calculate the working set of each user over time, we first extract the list of apps they launch, for every week period (Sunday to Saturday). Then, for each app in this users usage data, we calculate the number of weeks it appeared in the weekly working sets, and divide that by the total number of weeks that this user user remained in our dataset. In doing so, we calculate a probability $p$ of the app being on the each weekly working set. Note, for this analysis we filtered out data for users who were present in our dataset for a period less than 10 weeks, because a user who was present for too small number of weeks may skew our working set analysis.

\textbf{Working Set Size of Users:}  We depicted the number of apps that appeared with a probability higher than $p=0.9, 0.95, 1.0$ in Figure \ref{figure:user_working_set}. For example, for $p=0.9$, the figure plots a CDF of the number of apps that are launched at least 90\% of the weeks the user is in our dataset as a function of the working set. A higher $p$ leads to a smaller working set, for example for $p=1.0$ the working set is defined as the list of apps that are used every week. `For this case of $p=1.0$, the figure shows that around 78.5 \% of users used less than or equal to 10 apps as their working set, and about 91.2 \% of users used less than or equal to 14 apps on iPhones. If we relax the constraint to $p=.95$ or $p=.90$, we have one or two more apps in the working set, which means the working set has been already sufficiently covered by $p=1.0$ criteria. \emph{We conclude that each user's weekly working set lies around 14 to 18, covering roughly more than 90\% of users working set.}
\section{Discussion}

We started our analysis with a question: \emph{does app usage change over time?} We organized our analysis in combinations of two different dimensions of granularity. In this section, we discuss the implications for researchers looking into mobile device user behavior, as well as app developers, in addition to providing recommendations based on our results.

\subsection{User's App Usage Changes over Time}

\subsubsection{Changes in Entire Population Level}

The key results of our paper come from the length of the dataset we used for analysis, and thus our ability to study longitudinal changes. Our results indicate that mobile app usage does change significantly over time, first observing a longitudinal variability in average mobile usage in terms of per-app usage and per-category usage. These results demonstrate that looking at a population during a short period misses important variables in how app usage changes, and that expanding the research/testing period for mobile app usage behavior might be beneficial.

\subsubsection{Changes in Individual User Level}

Our analysis of changes in individual app usage behavior month by month showed high ratios (\%) between the standard deviation of usage and mean usage, as well as wide variance in the standard deviation of usage by app category. Interestingly, this high variance over time was not evident when examining the results at the entire population level, since while there were some changes (notably between Games and Productivity apps), our results showed that population usage was fairly constant.  

The high average variance on a monthly basis suggests that results which analyze user behavior over a shorter period may not be indicative of long-term user behavior, necessitating longer studies. Additionally, the breakdown of average standard deviation available in Table \ref{tab:SingleVarariance} may prove useful to both researchers and app developers, depending upon the category of app usage being considered. A longer study may be appropriate when there is high standard deviation, while a shorter study may be sufficient for categories with relatively small standard deviations. Depending on the purpose and accuracy required of a study, extrapolating long-term conclusions from short-term studies will not yield accurate long-term results.

\subsection{Working Set of Apps \& Longitudinally Popular Apps}

Out analysis on top $N$ most used app list provides suggests that we have roughly 10\% $/sim$ 20\% of those $N$ apps stay in that top $N$ list for the whole period. This can potentially provide a useful reference for optimizing systems and infrastructures around mobile ecosystem, because it implies that they can consider caching the contents related to those popular apps.

Our analysis shows that users have a small working set of apps ($\sim$18 apps) they occasionally launch over the course of a week. This finding can be leveraged by engineers or designers of mobile systems, because knowing how large user's working set size is can help them determining how many apps they will place by default in app dashboards or shortcut panels on mobile user interfaces.

\subsection{Impact of Form Factor on Device Usage}

Our longitudinal app usage analysis revealed that people use iPads and iPhones differently. Even though this is not surprising, rather expected, we quantitatively substantiated this common belief using a fine-grained dataset at scale. We also provide a series of detailed analysis around the differences between the two form factors, which we believe will be a good reference for mobile researchers and developers. 

More importantly, one of the major findings is that iPads and iPhones show different responses to longitudinal changes. This implies that researchers should model device usage differently based on form factor in the longitudinal context. For example, a huge longitudinal growth of a certain app category usage on iPhone would not necessarily imply the same amount of impact on iPads. A study that aims to predict or investigate app usage over time should split its target population based on the form factor, since parameters derived from an aggregated population of iPads and iPhone users may lead to an in accurate results. Because Android devices have much more diversity in its form factors (e.g. they have a wider diversity in screen sizes and resolutions), similar studies on Android devices should be conducted more carefully to respond to this difference.
\section{Conclusion}

In this paper, we focused on exploring whether app usage has changed over time from various points of view. We evaluated the longitudinal effects on app usage as well as app category level usage. Furthermore, we explored whether app usage in the entire population level and individual user level changes over time.

In summary, we find that both app-level and app category-level usage pattern changes over time across the entire dataset, reflecting a longitudinal shift of users' demand on smart devices. Along with the global change in typical app usage, we also find that each individual's usage pattern also changes over time, having higher variability than that of a typical user in the dataset. In addition, we also observed that users keep a small set of weekly working set of apps. Finally, we find that there is a subset of longitudinally popular apps in top $N$ list and the list remains quite similar across time.

The mobile device world is constantly evolving and studying mobile device usage behaviors will be increasingly challenging as devices become more powerful. For example, background use of navigation and split-screen multitasking on larger displays allow users to use several apps simultaneously to perform unique tasks. The paper contributes to this growing research area, which will help shape the systems and data analysis techniques of the future. 
\section{Acknowledgement}

We thank our anonymous reviewers for the feedback and their helpful comments. This work was supported in part by National Science Foundation CSR-1526237, TWC-1564009 and the DARPA Brandeis Program. We would also like to acknowledge the Scott Institute at Carnegie Mellon University and Google for their various gifts supporting this research.

\bibliographystyle{ACM-Reference-Format}
\bibliography{main}

% 
% If your work has an appendix, this is the place to put it.
%\appendix



%Lorem ipsum dolor sit amet, consectetur adipiscing elit. Morbi malesuada, quam in pulvinar varius, metus nunc fermentum urna, id sollicitudin purus odio sit amet enim. Aliquam ullamcorper eu ipsum vel mollis. Curabitur quis dictum nisl. Phasellus vel semper risus, et lacinia dolor. Integer ultricies commodo sem nec semper. 


%Etiam commodo feugiat nisl pulvinar pellentesque. Etiam auctor sodales ligula, non varius nibh pulvinar semper. Suspendisse nec lectus non ipsum convallis congue hendrerit vitae sapien. Donec at laoreet eros. Vivamus non purus placerat, scelerisque diam eu, cursus ante. Etiam aliquam tortor auctor efficitur mattis. 


%Nam id fermentum dui. Suspendisse sagittis tortor a nulla mollis, in pulvinar ex pretium. Sed interdum orci quis metus euismod, et sagittis enim maximus. Vestibulum gravida massa ut %felis suscipit congue. Quisque mattis elit a risus ultrices commodo venenatis eget dui. Etiam sagittis eleifend elementum. 

%Nam interdum magna at lectus dignissim, ac dignissim lorem rhoncus. Maecenas eu arcu ac neque placerat aliquam. Nunc pulvinar massa et mattis lacinia.

\end{document}